## aDORe: a modular, standards-based Digital Object Repository

Herbert Van de Sompel, Jeroen Bekaert, Xiaoming Liu, Luda Balakireva, Thorsten Schwander
Los Alamos National Laboratory, Research Library
{herbertv, jbekaert, liu_x, ludab, schwander}@lanl.gov

## Abstract

This paper describes the aDORe repository architecture, designed and implemented for ingesting, storing, and accessing a vast collection of Digital Objects at the Research Library of the Los Alamos National Laboratory. The aDORe architecture is highly modular and standards-based. In the architecture, the MPEG-21 Digital Item Declaration Language is used as the XML-based format to represent Digital Objects that can consist of multiple datastreams as Open Archival Information System Archival Information Packages (OAIS AIPs). Through an ingestion process, these OAIS AIPs are stored in a multitude of autonomous repositories. A Repository Index keeps track of the creation and location of all the autonomous repositories, whereas an Identifier Locator registers in which autonomous repository a given Digital Object or OAIS AIP resides. A front-end to the complete environment – the OAI-PMH Federator – is introduced for requesting OAIS Dissemination Information Packages (OAIS DIPs). These OAIS DIPs can be the stored OAIS AIPs themselves, or transformations thereof. This front-end allows OAI-PMH harvesters to recurrently and selectively collect batches of OAIS DIPs from aDORe, and hence to create multiple, parallel services using the collected objects. Another front-end – the OpenURL Resolver – is introduced for requesting OAIS Result Sets. An OAIS Result Set is a dissemination of an individual Digital Object or of its constituent datastreams. Both front-ends make use of an MPEG-21 Digital Item Processing Engine to apply services to OAIS AIPs, Digital Objects, or constituent datastreams that were specified in a dissemination request.

## 1. Introduction

When compared to most academic and research libraries, the Research Library of the Los Alamos National Laboratory (LANL) follows a rather unique strategy with respect to providing access to digital scholarly information. The general trend in digital library services is to have users access externally hosted materials through third party services, federated through a locally hosted Web Portal. In order to be self-supporting with respect to mission-critical scholarly information, the LANL library acquires or licenses a vast collection of digital scholarly assets, hosts those assets locally, and makes them accessible through locally developed user services. The locally hosted assets include secondary data feeds from BIOSIS, Inspec, Thomson Scientific, and primary information feeds from major scholarly publishers such as The American Physical Society, The Institute of Physics Publishing, Elsevier, and Wiley.

At the time of writing the collection amounts to approximately 80,000,000 locally hosted assets. In many cases these assets are *complex* in the sense that they consist of multiple individual datastreams that jointly form a single logical unit. That logical unit can, for example, be a scholarly publication that consists of:
- A research paper in both PDF and ASCII format,
- Metadata describing the paper and references made in the paper expressed in XML,
- Auxiliary datastreams such as images and videos in various formats, such as TIFF, JPEG and MPEG.

For reasons of clarity, this paper will refer to an asset as a *Digital Object*, and to the individual datastreams of which the asset consists as *constituent datastreams*.

Hosting, archiving and making accessible such a vast and heterogeneous collection of Digital Objects in a consistent and sustainable manner is a challenge that touches on many areas of digital library practice and research, including the identification of Digital Objects and constituent datastreams, the expression of relationships between Digital Objects, the representation of Digital Objects by means of object models, and methods to ingest, store, and access Digital Objects.

Over the last 2 years, the Digital Library Research and Prototyping Team of the LANL Research Library has worked on the design of the aDORe repository architecture aimed at ingesting, storing, and making accessible to downstream applications an ever growing heterogeneous digital collection. A first version of the design has been implemented and been brought into production, while work is currently ongoing to implement a more robust second version.

While the aDORe design was inspired by requirements imposed by the LANL Research Library, the authors feel that the architecture has properties that should be attractive for other repository projects:

- Standards-based design: Throughout the architecture, standards or defacto standards are used. These include W3C XML, W3C XML Schema, the MPEG-21 Digital Item Declaration (MPEG-21 DID), the MPEG-21 Digital Item Identification (MPEG-21 DII), the MPEG-21 Digital Item Processing (MPEG-21 DIP), the Open Archives Protocol for Metadata Harvesting (OAI-PMH), the NISO OpenURL Framework for Context-Sensitive Services, and the Internet Archive ARC file format.
- Natively component-based, distributed design: The architecture operates on the basis of various autonomous components; interaction with those components is protocol-based.
- The dynamic binding of dissemination methods to stored Digital Objects and constituent datastreams.

Figure 1 introduces the major components of the aDORe architecture. All components are explained in detail in the remainder of this paper; a summary is provided here:
The ingestion process is represented at the extreme right hand side, while downstream applications that interact with the repository are situated at the extreme left hand side.

- At the right hand side, Figure 1 shows a multitude of *Autonomous OAI-PMH Repositories* [21]. Each of these autonomous repositories stores a collection of Open Archival Information System [15] Archival Information Packages (OAIS AIPs) each of which represents a Digital Object according to the MPEG-21 DID specification [17,43]. Section 2 is dedicated to detailing aspects related to these Autonomous OAI-PMH Repositories.
- The *Repository Index* is shown below the Autonomous OAI-PMH Repositories. It is a registry that keeps track of the creation and location of Autonomous OAI-PMH Repositories in the aDORe environment. This component, which is also accessible through the OAI-PMH [21], is detailed in Section 3.
- At the left center of Figure 1 is the *Identifier Locator*. For each OAIS AIP stored in aDORe, this component contains the identifiers associated with the OAIS AIP itself and with the Digital Object it represents. It also contains the location of the Autonomous OAI-PMH Repository in which the OAIS AIP and hence the Digital Object reside. When multiple versions of the same Digital Object exist, the Identifier Locator keeps track of all locations. The Identifier Locator can be populated through batch loading or OAI-PMH harvesting. It can be queried in a variety of ways, including the Handle protocol [36]. This component is described in Section 4.
- To the right of the Identifier Locator, Figure 1 shows the server-side *MPEG-21 DIP Engine* and its associated components (DIBO/DIXO registry, DIM Inserter, DIP Table). These components are introduced to facilitate the disseminations of stored OAIS AIPs, Digital Objects, and their constituent datastreams. Section 5 provides insights in their operation.

- At the top left, Figure 1 shows the *OAI-PMH Federator*. This components turns the complete aDORe environment into a single OAI-PMH repository, thereby hiding all architectural details and complexities from downstream harvesters. The OAI-PMH Federator becomes the single point of access for off-the-shelf OAI-PMH harvesters that aim to collect batches of OAIS AIPs from the aDORe environment. The OAI-PMH Federator interacts with other components of the environment mainly using the OAI-PMH. It calls upon the MPEG-21 DIP Engine when transformations of stored OAIS AIPs are requested, rather than the stored OAIS AIPs themselves. Details are provided in Section 6.
- Finally, at the bottom left, Figure 1 shows the *OpenURL Resolver*. This component provides a front-end to the aDORe environment from which various disseminations of an individual Digital Object or its constituent datastreams can be obtained using requests that are compliant with the NISO OpenURL standard [30]. The OpenURL Resolver interacts with other components of the environment mainly using the OAI-PMH. It calls upon the MPEG-21 DIP Engine to deliver the requested disseminations. The OpenURL Resolver is also described in Section 6.

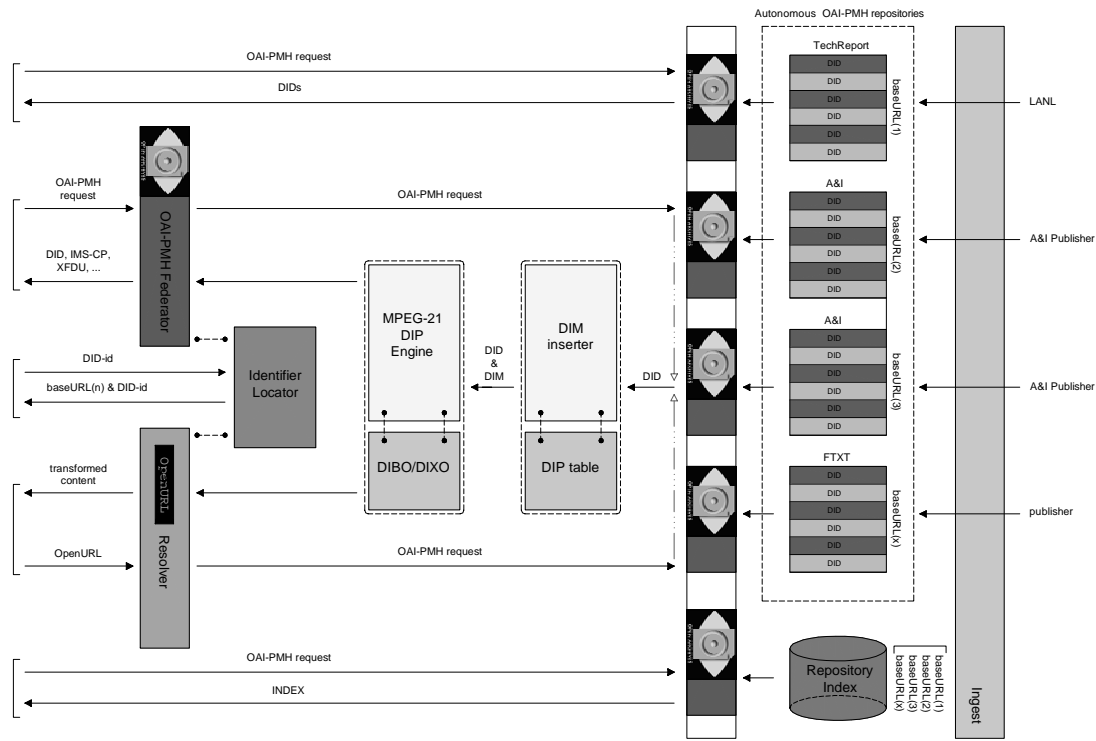

**Figure 1.** The aDORe Architecture

## 2. Ingesting assets into the aDORe environment

### 2.1. Representing Digital Objects

The complex nature of the assets to be ingested into aDORe led to an investigation regarding existing approaches to wrap constituent datastreams into a single wrapper structure that could function as an Open Archival Information System [15] Archival Information Package (OAIS AIP). This quickly led to an interest in representing assets by means of XML wrappers, which itself resulted in the selection of the MPEG-21 Digital Item Declaration (MPEG-21 DID) [17,43] as the sole way to represent assets as Digital Objects in aDORe. Several reasons motivated the choice of MPEG-21 DID over competing specifications. As opposed to some other specifications, MPEG-21 DID defines a data model, which can be instantiated using various technologies. Currently, both a binary instantiation, and an instantiation based on XML Schema exists, but one could imagine the definition of, for example, an RDF-based instantiation in the future. This approach allows for implementing solutions across various and evolving technical environments while maintaining basic architectural concepts. Also, MPEG-21 DID takes an approach that enforces cross-community interoperability, while allowing the flexibility for the emergence of compliant, community-specific profiles. The XML Schema instantiation is surprisingly elegant and simple, making the development of tools quite straightforward. Furthermore, the combination of MPEG-21 DID with the MPEG-21 Digital Item Identification (MPEG-21 DII) results in an unambiguous way to handle identifiers, a core feature that is remarkably absent in other specifications. And, importantly, MPEG-21 DID is an ISO standard developed by major players in the content and technology industry, which provides some guarantees regarding its adoption, and the emergence of compliant tools.

MPEG-21 DID introduces a set of abstract concepts that, together, form a well-defined Abstract Model for declaring Digital Objects. Based on those abstract concepts, MPEG-21 DID defines the Digital Item Declaration Language (MPEG-21 DIDL), an XML representation of the MPEG-21 DID Abstract Model that provides broad flexibility and extensibility for the XML-based representation of Digital Objects. A Digital Object represented according to the MPEG-21 DIDL XML syntax is called a DIDL document. Interested readers are referred to [4] for more information on MPEG-21 DID. A simplified explanation of the MPEG-21 DID Abstract Model is given below; it is also illustrated in Figure 2. The model recognizes several MPEG-21 DID entities (written in *italic* font style), each of which has a corresponding XML element in the DIDL XML Schema [43]. The `courier` font is used to refer to XML elements.

- A *container* is a grouping of *containers* and/or *items*.
- An *item* is the declarative representation of a Digital Object. It is a grouping of *items* and/or *components*.
- A *component* is a grouping of *resources*. Multiple *resources* in the same *component* are considered bit-equivalent and consequently it is left to an agent to select which one to use.
- A *resource* is an individual datastream.
- Secondary information pertaining to a *container*, an *item*, or a *component* can be conveyed by means of a *descriptor/statement* construct.

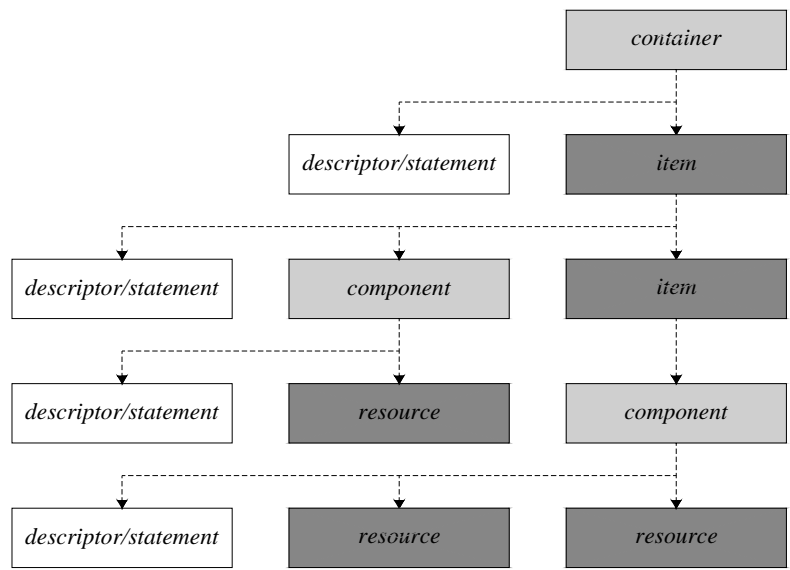

**Figure 2.** Example of a document structure conformant with the MPEG-21 DID specification

At LANL, Digital Objects to be stored in aDORe can, in principle, be obtained in a variety of ways including FTP, OAI-PMH resource harvesting [42], Web crawling and delivery on physical media. An ingestion process has been developed that represents each Digital Object according to the MPEG-21 DID specification. Hereby, a DIDL XML document is created that functions as the OAIS AIP that represents the Digital Object. This OAIS AIP is generated as follows:

- Each Digital Object is mapped to a top-level DIDL `Item` element. Constituent datastreams are provided in child elements of this top-level `Item`. An identifier of the Digital Object is expected at this level.
- A constituent datastream of a Digital Object is provided in a `Component/Resource` construct. Such constructs are themselves embedded in a sub-`Item` of the top-level `Item`, or in the top-level `Item` itself, depending on whether or not they have identifiers in their own right.
- For pragmatic reasons, explained in Section 6, related to dissemination of OAIS AIPs, the top-level `Item` is embedded in a `Container` element.
- The top-level `Item` and its parent `Container` element are then embedded in the `DIDL` root element to obtain a DIDL XML document that is the OAIS AIP that represents the Digital Object.

An important, OAIS-inspired, characteristic of the aDORe environment is that whenever a new version of a previously ingested Digital Object needs to be ingested, a new DIDL document is created for it. Existing DIDL documents are never updated or edited. A new version of a Digital Object may, for example, become available because an information provider delivers an updated copy, or because a digital preservation strategy requires the migration of a file format used in the Digital Object. As will be shown, the Identifier Locator keeps track of all versions of a Digital Object.

A DIDL document can physically contain or reference the constituent datastreams of the Digital Object that it represents. The DIDL document also contains identifiers, as well as information such as collection membership, XML Signature constructs, media format information pertaining to constituent datastreams, and provenance information. The actual use of the MPEG-21 DIDL in the LANL Repository, and the LANL DIDL profile is described in some detail in the slightly outdated [1] and the more recent [4].

Annex A shows the DIDL document that represents the sample Digital Object – a scholarly paper – as shown in Table 1.

|  | Type | MIME | identifier |
|---|---|---|---|
| **Digital Object** | scholarly paper | N/A | DOI |
| **Constituent Datastream 1** | metadata record | application/xml | PMID |
| **Constituent Datastream 2** | fulltext file | application/pdf | – |

**Table 1.** Building blocks of the sample Digital Object

## 2.2. Identifying DIDL documents, Digital Objects and constituent datastreams

A good understanding of the meaning and use of identifiers in the aDORe environment is crucial to understand its very design. aDORe uses two parallel identification mechanisms that are described below. Figure 3 illustrates the identifier approach by means of our sample Digital Object:

- ***Content Identifiers***. Content Identifiers corresponds to what the OAIS [15] categorizes as Content Information Identifiers. Content Identifiers are directly related to identifiers that are natively attached to Digital Objects before their ingestion into aDORe. In many cases such Digital Objects, or their constituent datastreams, have identifiers that were associated with them when they were created or published. Content Identifiers in aDORe are expressed as URIs. For the above sample Digital Object, the following Content Identifiers exist:
  - The Digital Object has a Digital Object Identifier [21]. This is expressed as a URI using the proposed info URI scheme [40] to obtain the Content Identifier 'info:doi/10.123/44455'.
  - The metadata record that describes the paper has a PubMed identifier. This is expressed as a URI using the proposed info URI scheme to obtain the Content Identifier 'info:pmid/2225887'.
  - The paper in PDF format itself does not have a Content Identifier.

  During the ingestion process, a `Descriptor/Statement` construct attached to the top-level `Item` is used to convey the Content Identifier of the Digital Object, using the `Identifier` element of the MPEG-21 DII XML Namespace [3,18]. This can be seen in the sample Digital Object of Annex A, where the top-level `Item` has Content Identifier 'info:doi/10.123/44455'. As previously mentioned, the ingestion process will create sub-`Items` for all constituent datastreams of a Digital Object that have Content Identifiers. This can be seen in the sample Digital Object of Annex A, where a sub-`Item` contains the metadata that describes the scholarly paper; the Content Identifier of the metadata – 'info:pmid/2225887' – is again conveyed using the `Identifier` element of the MPEG-21 DII XML Namespace. Because the PDF file does not have a Content Identifier, the ingestion process maps it to a `Component` element for which no `Identifier` element is provided. That `Component` element is a child of the top-level `Item`.

- ***Package Identifiers.*** A DIDL document that represents a Digital Object functions as an OAIS AIP in aDORe. During the ingestion process, this DIDL document itself is accorded a globally unique identifier, which the OAIS categorizes as an Archival Information Package Identifier. This Package Identifier is conveyed by means of the `DIDid` attribute from the LANL DIEXT Namespace, attached to the `DIDL` root element. Values are constructed using the UUID algorithm [22]; they are expressed as URIs in a reserved sub-namespaces of the 'info:lanl-repo/' namespace, which the LANL Research Library has registered under the info URI Scheme [40]. Also, during the ingestion process, `Container`, `Item` and `Component` elements receive globally unique XML IDs, again created using the UUID algorithm. As a result, these XML elements become globally addressable using a combination of the Package Identifier of the DIDL document in which they are contained, and their own XML ID. In our sample Digital Object, this addressing mechanism is as follows (for readability, UUID values have been shortened):
  - The Package Identifier of the DIDL document which represents the Digital Object: info:lanl-repo/i/58f202ac
  - The `Item` that represent the Digital Object has XML ID 'uuid-00005e90'. As a result, it can be addressed as 'info:lanl-repo/i/58f202ac#uuid-00005e90'. As was shown, using Content Identifiers, it can also be addressed as 'info:doi/10.123/44455'.
  - The sub-`Item` for the constituent metadata datastream of the Digital Object has XML ID 'uuid-8881b35e'. As a result, it can be addressed as 'info:lanl-repo/i/58f202ac#uuid-8881b35e' [11]. As was shown, using Content Identifiers, it can also be addressed as 'info:pmid/2225887'.
  - The `Component` that contains the paper in PDF format has XML ID 'uuid-00004a42'. Hence, it can be addressed as 'info:lanl-repo/i/58f202ac# uuid-00004a42'. It can not be addressed using a Content Identifier.

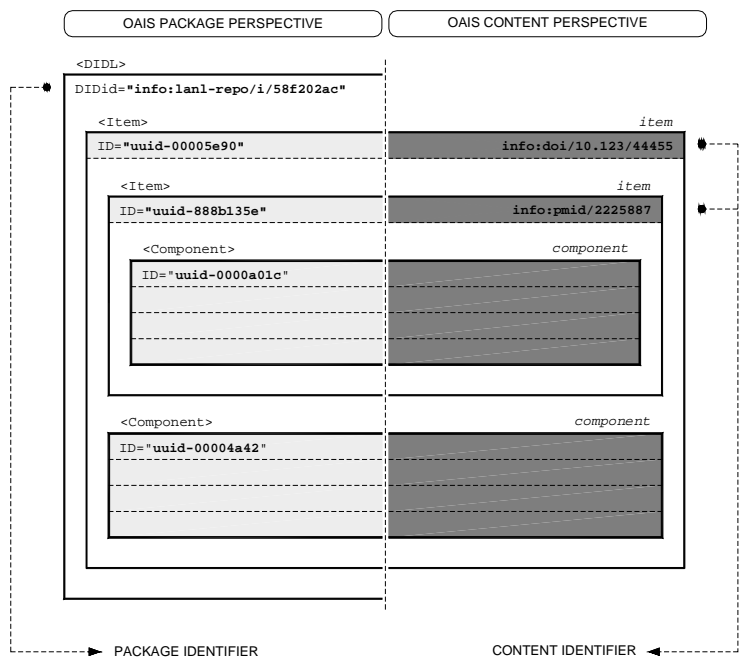

**Figure 3.** Content Identifiers and Package Identifiers provide parallel identification mechanisms

## 2.3. Storing Digital Objects in Autonomous OAI-PMH Repositories and ARC files

The ingestion process turns each delivered Digital Object into a DIDL document. Typically, these DIDL documents physically contain constituent XML-based metadata datastreams of the Digital Object inline. To keep the size of the DIDL documents small, and hence to make them easier to process, other constituent datastreams of the Digital Object are provided by reference and are physically stored in Internet Archive ARC files [5]. Each DIDL document itself is stored in one of many Autonomous OAI-PMH Repositories that exist in the aDORe environment.

The Open Archives Protocol for Metadata Harvesting (OAI-PMH) is a low-barrier, HTTP-based protocol. It has been specified to allow incremental harvesting of XML metadata. An OAI-PMH repository is a network accessible server that can process the 6 OAI-PMH protocol requests, and respond to them as specified by the protocol document. A harvester is an application that issues OAI-PMH protocol requests, in order to harvest XML metadata. The OAI-PMH is based on a data model that helps specifying the semantics of the 6 protocol requests. That data model is depicted in Figure 4, for cases where the resource about which an OAI-PMH repository exposes metadata, is digital content. In what follows, OAI-PMH entities are written in *italic*, while OAI-PMH protocol requests are written in `courier`.

- At the very top is a digital *resource* (a PDF file, for example) about which an OAI-PMH repository exposes *metadata*. Note that the digital *resource* can also be compound, i.e. consist of multiple datastreams. By definition, *resources* themselves are outside of the scope of the OAI-PMH.
- Listed below the *resource* is the *item*. The *item* is the highest-level entity within the scope of the OAI-PMH. In essence, the *item* is the entry point to all available *metadata* pertaining to a *resource*. In the protocol, the *item* is uniquely identified by an OAI-PMH *identifier*.
- Below the *item*, several *records* are shown. *Records* contain *metadata* (and secondary information about that *metadata*). A specific *record* in the OAI-PMH is unambiguously identified by means of the combination of the OAI-PMH *identifier* (of the *item*), the *metadataPrefix* that specifies the *metadata format* used for the dissemination of the *metadata*, and the OAI-PMH *datestamp* of the *metadata*. The *datestamp* is the date and time of creation or modification of *metadata*. Note that the *datestamp* is a property of the *metadata* record, not of the *item* as used to be the case in previous protocol versions [21]. This reflects the fact that *metadata* of various *metadata formats* may be made available and may be modified independently, thus having different *datestamps*.

- The OAI-PMH also defines a *set* – not depicted in Figure 4 – as an optional construct for grouping *items* for the purpose of selective harvesting. Repositories may organize *items* into *sets*. *set* organization may be flat, i.e. a simple list, or hierarchical. Multiple, parallel, *set* structures may exist.

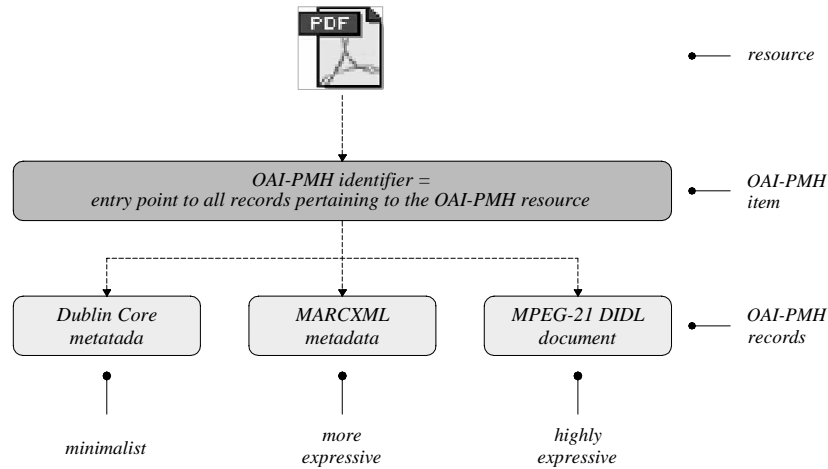

**Figure 4.** The OAI-PMH data model

The OAI-PMH defines 3 supporting protocol requests that are aimed at helping a harvester understand the nature of an OAI-PMH Repository:

- `Identify`: this verb is used to retrieve information about a repository; an important information element returned in the response to the `Identify` request is the granularity of the datestamp supported by the repository (day-level or seconds-level).
- `ListMetadataFormats`: this verb is used to retrieve the *metadata formats* available from a repository.
- `ListSets`: This verb is used to retrieve the *set* structure of a repository. This information is useful for selective harvesting.

The OAI-PMH defines 3 further protocol requests that are aimed at the actual harvesting of XML metadata:

- `ListRecords`: this verb is used to harvest *records* from a repository. Optional arguments permit selective harvesting of *records* based on *set* membership and/or *datestamp*.
- `GetRecord`: This verb is used to retrieve an individual *record* from a repository. Required arguments specify the *identifier* of the *item* from which the *record* is requested and the *metadata format* of the metadata that should be included in the *record*.
- `ListIdentifiers`: This verb is an abbreviated form of `ListRecords`, retrieving only *identifiers*, *datestamps* and *set* information.

Due to its origins in the realm of resource discovery, the OAI-PMH mandates the support of the Dublin Core [30] metadata format, but strongly encourages supporting more expressive formats.

As a result, any metadata format can be used as long as it is defined by means of an XML Schema [10]. In typical use cases, the exposed *metadata* is descriptive, and is expressed by means of *metadata formats* of varying complexity, such as simple Dublin Core, or MARCXML [24]. The Autonomous OAI-PMH Repositories in aDORe, however, support metadata that are highly expressive and accurate in their representation of Digital Objects by using MPEG-21 DIDL as a metadata format, and hence by exposing DIDL documents as metadata to harvesters.

Each such Autonomous OAI-PMH Repository has the following characteristics:
- It has a unique, persistent `baseURL`, the http address `BaseURL(n)`.
- Contained `records` are DIDL documents only.
- The `identifier` used by the OAI-PMH is the Package Identifier.
- The `datestamp` used by the OAI-PMH is the creation DateTime of the DIDL document.
- The only supported metadata format is DIDL, with `metadataPrefix` DIDL, defined by the MPEG-21 DIDL XML Schema.
- The supported OAI-PMH harvesting `granularity` is seconds-level.
- `Set` structures may be supported, but to reduce complexity this aspect will not be discussed in this paper.

As a result, harvesters operated by downstream service providers – such as indexing engines – can use the selective harvesting capabilities of the OAI-PMH (`datestamp` and `set`) to recurrently collect OAIS AIPs from the Autonomous OAI-PMH Repositories. Because OAIS AIPs are never updated, incremental harvesting will yield newly added OAIS AIPs only.

At LANL, information providers typically deliver sizable batches of additions or updates on a periodic basis. In those cases, an Autonomous OAI-PMH Repository is created per delivered batch, and a special-purpose storage mechanism – the XMLtape - is used. An XMLtape is a valid and well-formed XML file that concatenates a large batch of DIDL documents. Special-purpose software has been created that turns an XMLtape into an OAI-PMH repository [44]. The usage of the combination of XMLtapes and Internet Archive ARC files as a means to store ingested Digital Objects is described in [37].

## 3. A registry of Autonomous OAI-PMH Repositories: The Repository Index

As has been shown, updates can be harvested from the Autonomous OAI-PMH Repositories. However, the question remains unanswered as to how harvesters learn about the existence, addition of, and location of these Autonomous OAI-PMH Repositories. In order to provide this crucial intelligence, the *Repository Index* is introduced.

The Repository Index contains an entry for each Autonomous OAI-PMH Repository in the aDORe environment, containing the following information:
- The repository baseURL: The `baseURL` of an Autonomous OAI-PMH Repository, which is a unique and persistent URI.
- The repository creation DateTime: The time when the Autonomous OAI-PMH Repository became harvestable, by its very appearance in the Repository Index. This time is expressed as an ISO 8601 [14] DateTime with seconds granularity.
- Metadata pertaining to the creation of the Autonomous OAI-PMH Repository, the nature of its content, ...

It cannot be overlooked that the first two information elements of the Repository Index map directly to the notions of the `identifier` and the `datestamp` of the OAI-PMH, respectively. And, indeed, in aDORe, the Repository Index is exposed as an OAI-PMH repository in its own right, with the following properties:

- It has a unique, persistent `baseURL`, the http address `BaseURL(Repo-Index)`.
- Contained records comply with a locally defined XML-based metadata format, identified by `metadataPrefix` INDEX, which facilitates the expression of the necessary metadata about Autonomous OAI-PMH Repositories.
- The `identifier` used by the OAI-PMH is the repository-baseURL `BaseURL(n)`.
- The `datestamp` used by the OAI-PMH is the repository creation DateTime. There are no updates to metadata contained in the Repository Index, and hence this `datestamp` will never change and always remain equal to the time the Autonomous OAI-PMH Repository became available for harvesting in the aDORe environment.
- The supported OAI-PMH harvesting `granularity` is seconds-level.
- `set` structures may be supported, but to reduce complexity, this aspect will not be discussed in this paper. Typically, `set` structures would be used in the Repository Index to broadly categorize the nature or content of Autonomous OAI-PMH Repositories.

## 4. Locating DIDL documents, Digital Objects, and constituent datastreams: The Identifier Locator

Harvesters working on behalf of service providers collect DIDL documents from the aDORe environment, and build services with the contained Digital Objects. As a result, identifiers contained in the harvested DIDL documents become available in applications such as search engines. As explained in Section 2.2, these identifiers can either be Package Identifiers identifying DIDL documents or contained DIDL XML elements, or Content Identifiers identifying Digital Objects or constituent datastreams. It is essential that, when such identifiers show up in downstream applications, the identified resource can be retrieved from the aDORe environment. For this purpose the *Identifier Locator* is introduced. The Identifier Locator collects the following identifier-related information from the aDORe environment:

- For OAIS AIPs: The Package Identifier of each OAIS AIP in the aDORe environment, and for each Package Identifier the `baseURL` of the Autonomous OAI-PMH Repository in which the OAIS AIP with the given Package Identifier resides. In Table 2, 'info:lanl-repo/i/58f202ac' and 'info:lanl-repo/i/002035b2' are Package Identifiers of DIDL documents located in the OAI-PMH repository with `baseURL` 'BaseURL(3)' and 'BaseURL(6)', respectively. The first entry of Table 2 is obtained from our sample DIDL document.
- For Digital Objects and their constituent datastreams: The Content Identifier of each Digital Object and constituent datastream stored in the aDORe environment. For each such Content Identifier, the Package Identifiers (including XML ID) of all OAIS AIPs that contain a Digital Object or constituent datastream with the given Content Identifier. In Table 3, Content Identifiers 'info:doi/10.123/44455' and 'info:lanl-repo/biosis/abcdef' identify resources contained in the DIDL documents with Package Identifier 'info:lanl-repo/i/58f202ac' and 'info:lanl-repo/i/002035b2', respectively. Within these DIDL documents, the XML elements that contain resources with the aforementioned Content Identifiers have XML IDs 'uuid-00005e90', and 'uuid-00007y55', respectively. Table 3 also shows that the resource with Content Identifier 'info:pmid/2225887' is represented by two OAIS AIPs with Package Identifiers 'info:lanl-repo/i/58f202ac' and 'info:lanl-repo/i/12e303be', respectively, indicating the existence of different versions. As can be seen, the first two entries in Table 3 are obtained from our sample DIDL document.

| Package Identifier | OAI-PMH repository |
|---|---|
| info:lanl-repo/i/58f202ac | BaseURL(3) |

| info:lanl-repo/i/002035b2 | BaseURL(6) |
|---|---|

**Table 2.** Package Identifier module of the Identifier Locator

| Content Identifier | Package Identifier | XML ID |
|---|---|---|
| info:doi/10.123/44455 | info:lanl-repo/i/58f202ac | uuid-00005e90 |
| info:pmid/2225887 | info:lanl-repo/i/58f202ac | uuid-8881b35e |
| info:lanl-repo/biosis/abcdef | info:lanl-repo/i/002035b2 | uuid-00007y55 |
| info:pmid/2225887 | info:lanl-repo/i/12e303be | uuid-875646ae |

**Table 3.** Content Identifier module of the Identifier Locator

By default, the Identifier Locator is populated through recurrent OAI-PMH harvesting from the Autonomous OAI-PMH Repositories, but for cases that involve massive ingestion of Digital Objects, a faster batch loading mechanism has been implemented. The Identifier Locator is accessible to downstream applications via the Handle protocol [36]; an SRW [27] SOAP-based [12] interface will be implemented later. Through consultation of the Identifier Locator, an application can obtain the information necessary to use the OAI-PMH to retrieve the DIDL document with a specified Package Identifier, or the DIDL document in which a resource with a specified Content Identifier resides. This works as follows:

- If the resource with identifier 'info:lanl-repo/i/58f202ac' is requested, a look-up in the Identifier Locator will learn that it is a DIDL document with Package Identifier 'info:lanl-repo/i/58f202ac' that is located at 'BaseURL(3)'. This DIDL document can be obtained by issuing the OAI-PMH request `[BaseURL(3)?verb=GetRecord&identifier=info:lanl-repo/i/58f202ac&metadataPrefix=DIDL]`

- If the resource with identifier 'info:lanl-repo/i/58f202ac#uuid-00005e90' is requested, a look-up in the Identifier Locator will learn that the identified resource is an XML element contained in the DIDL document with Package Identifier 'info:lanl-repo/i/58f202ac'. Furthermore, the lookup will learn that this DIDL document is located at 'BaseURL(3)'. Again, that DIDL document can be obtained by issuing the OAI-PMH request `[BaseURL(3)?verb=GetRecord&identifier=info:lanl-repo/i/58f202ac &metadataPrefix=DIDL]`, and from the resulting DIDL document, the XML element with XML ID 'uuid-00005e90' can be extracted.

- If the resource with identifier 'info:lanl-repo/biosis/abcdef' is requested, a look-up in the Identifier Locator will learn that it is available from a DIDL document with Package Identifier 'info:lanl-repo/i/002035b2', and that – in this DIDL document – it has the XML ID 'uuid-00007y55'. A look-up of this Package Identifier learns that the corresponding DIDL document is located at 'BaseURL(6)'. This DIDL document can be obtained by issuing the OAI-PMH request `[BaseURL(6)?verb=GetRecord&identifier=info:lanl-repo/i/002035b2&metadataPrefix=DIDL]`, and from the resulting DIDL document, the XML element with XML ID 'uuid-00007y55' can be extracted. This approach allows for a mapping of Content Identifiers to Package Identifiers.

## 5. Applying services to DIDL documents, Digital Objects and constituent datastreams: The MPEG-21 Digital Item Processing Engine

In order to facilitate delivery of various disseminations of stored DIDL documents, Digital Objects, and constituent datastreams, a separate component is introduced in the aDORe environment. This component – the *Digital Item Processing Engine* - operates according to the current version of the MPEG-21 Digital Item Processing (MPEG-21 DIP) specification [8], which remains to be standardized. By MPEG-21 definition, an MPEG-21 DIP Engine is capable of applying services to a Digital Object, or to its constituent datastreams. At the time of writing, the MEPG-21 DIP specification does not detail how to apply services to a complete DIDL document. Because the aDORe environment requires this capability, a pragmatic solution – described in Section 5.2. – has been introduced, awaiting a standardized solution from a future version or amendment on the MPEG-21 DIP specification.

## 5.1 Conveying processing information: MPEG-21 Digital Item Methods

MPEG-21 Digital Item Processing (MPEG-21 DIP) specifies an architecture pertaining to the dissemination of the *container, item, and component* entities of the MPEG-21 DID Abstract Model. To that end, this tenth MPEG-21 Part introduces the concept of a *Digital Item Method* (DIM), which is conceptually closely related to Fedora's 'behavior' concept [34,35]. A DIM is physically contained in the same DIDL document as the entity of the MPEG-21 DID Abstract Model with which it is associated. In the current, pre-standard, practice a DIM is typically accommodated in a `Component` element which must contain at least 2 other child constructs:

- A `Resource` element that contains or references the actual method. In MPEG-21 DIP, the DIM code is expressed using a language that is largely based on the ECMAScript Language [16]. As will be explained in the section on the MPEG-21 DIP Engine, this ECMAScript does not contain all the actual code required to implement the service. Rather, the ECMAScript is used to bootstrap and coordinate the implementation of the service. It contains calls to operations that do the real work.
- A `Descriptor/Statement` construct that contains a sequence of `Argument` elements from the MPEG-21 DIP XML Namespace, one per argument of the DIM. The value of an `Argument` element always identifies an XML element of a DIDL document.

A DIM is associated with an XML element of a DIDL document using the `objectType/Argument` technique, also explained in the slightly dated [1]:

- Using a special-purpose `Descriptor/Statement` construct, an MPEG-21 DID entity (`Container`, `Item`, `Component`) can be accorded an `ObjectType` element from the MPEG-21 DIP XML Namespace.
- Using another special-purpose `Descriptor/Statement` construct, a DIM can be accorded an `Argument` element, also from the MPEG-21 DIP XML Namespace.
- When, in a DIDL document, the value of an `ObjectType` element of an MPEG-21 DID entity (`Container`, `Item`, `Component`), and the value of the `Argument` of a DIM are equal, then the DIM can be applied to that MPEG-21 DID entity.

This approach is highlighted in the DIDL document shown in Annex B, in which a service is associated with a `Component` through the value 'urn:uuid:8f64eabf'. That value is used as the content of both the `ObjectType` element of the `Component`, and the `Argument` element of the DIM. The MPEG-21 DIP specification allows entities to have more than one `ObjectType`. Also, a DIM can bind to more than one entity by using multiple `Argument` elements, each of which connects via the `ObjectType` to the entities.

## 5.2 Dynamically associating services with DIDL documents, Digital Objects and constituent datastreams: Placeholders and DIP Table

As can be understood from the above, the MPEG-21 DIP Engine introduces the capability to deliver various disseminations of a stored Digital Object and of its constituent datastreams in the aDORe environment. Unfortunately, at the time of writing, the MPEG-21 DIP Engine specification does not pertain to services applied to a DIDL document itself. Such services are important in the aDORe environment to allow the dissemination of transformations of stored DIDL documents.

Transformations of practical use include the mapping of DIDL to other complex object formats such as METS [25] and IMS-CP [13], and the transformation of stored DIDL documents to XML documents that contain identifier-related information only, which can readily be used to populate the Identifier Locator. Awaiting the inclusion in the MPEG-21 DIP specification of capabilities to apply services to complete DIDL documents, a pragmatic solution has been introduced in aDORe that consists of conveying all such services at the level of a `container` element that is artificially included in all DIDL documents during ingestion.

In aDORe, DIMs are not embedded in the DIDL documents that are stored in the Autonomous OAI-PMH Repositories. Embedding DIMs would yield a need to frequently update the stored DIDL documents to add new or to edit existing DIMs as new processing methods emerge, or as existing ones are updated. Given the anticipated size of aDORe, the administrative overhead in doing so would be extensive if not forbidding. Also, the need to regularly 'touch' stored DIDL documents is a poor fit with the rather static nature of the content currently stored in aDORe, and with the previously described strategy to create new DIDL documents – instead of updating existing ones – when some form of editing has been performed. Therefore, aDORe handles the inclusion of DIMs dynamically, based on the content of DIDL documents. In the current implementation of this approach, 'placeholders' for DIMs, instead of actual DIMs, are embedded in stored DIDL documents during the ingestion process. When the dissemination of a stored DIDL document is requested, the *DIM Inserter* module dynamically adds DIMs to that DIDL document in a process that is based on the values of the placeholders contained in the DIDL document. The result is a so-called Completed DIDL document (see Annex B), which contains all resources that are present in the stored DIDL document, as well as all associated methods. It is worthwhile noting that the DIM Inserter to some extent resembles the 'context broker' mechanism described in [9].

Placeholder values are conveyed in DIDL documents via a `format` element from the Dublin Core Element Set [29]. Placeholders can be attached at the *container*, *item*, and *component* level of a DIDL document:

- As described earlier, at the *container* level, the placeholder is provided to allow the association of methods aimed at transforming complete DIDL documents. The placeholder value is the same for all stored DIDL documents. Again, it should be noted that passing on information pertaining to the DIDL document via *descriptor/statement* constructs at the *container* level is not compliant with the MPEG-21 DID model. As such, this application-specific *container*-based solution will be deprecated as soon as MPEG-21 DIP standardizes a mechanism to associate services with complete DIDL documents.
- At the *item* level, the placeholder conveys to which 'family' the DIDL document belongs. 'Family' information is closely related to the ingestion process. For example, the placeholder value 'info:lanl-repo/pro/ai' shown in Table 3 is used to convey that the Digital Object represented by a DIDL document is a descriptive metadata record.
- At the *component* level, the placeholder specifies a globally unique identifier for the digital format of the contained datastream. Currently, local identifiers are used that closely relate to MIME content types, but, once established, identifiers from the Global Format Registry [1] could be used. An example is shown in the second entry of Table 4.

The actual insertion of DIMs in a DIDL document is achieved through a look-up in a special-purpose registry – the *DIP Table*. The DIP Table lists all services that can be associated with DIDL documents, Digital Objects and contained datastreams stored in aDORe. In the DIP Table, each service has a Service Identifier. Also, for each service, the DIP Table contains a placeholder value – as used in the DIDL documents – with which the service is associated. For example, the first 2 entries of Table 4 shows that the service with Service Identifier 'info:lanl-repo/service/' is associated with all MPEG-21 DID entities that have 'info:lanl-repo/pro/ai' or 'info:lanl-repo/pro/paper' as a placeholder value. For each service, the DIP Table also lists a pointer to the DIM code that actually implements it.

| Service Identifier | info:lanl-repo/service/ |
|---|---|
| placeholder value | info:lanl-repo/pro/ai |
| Pointer to DIM code | http://purl.lanl.gov/dip/methods/toc.js |
| *Description* | *Service that displays a Table of Contents of a Digital Object as an XHTML page, listing all constituent datastreams as well as the services that are available for them.* |

| Service Identifier | info:lanl-repo/service/ |
|---|---|
| placeholder value | info:lanl-repo/pro/paper |
| Pointer to DIM code | http://purl.lanl.gov/dip/methods/toc.js |
| *Description* | *Service that displays a Table of Contents of a Digital Object as an XHTML page, listing all constituent datastreams as well as the services that are available for them.* |

| Service Identifier | info:lanl-repo/service/marc_2_mods |
|---|---|
| placeholder value | info:lanl-repo/fmt/3 |
| Pointer to DIM code | http://purl.lanl.gov/dip/methods/marctomods.js |
| *Description* | *Service that disseminates a stored MARCXML datastream as a MODS datastream* |

**Table 4.** Three entries of the DIP Table

The dynamic insertion of DIMs in a DIDL document, as performed by the DIM Inserter module is explained by means of the examples provided in Annex A and B. Consider the retrieval of the DIDL document that represents our sample Digital Object from aDORe (see Annex A), and focus on the MARCXML [18] metadata datastream. This datastream has a placeholder with a value of 'info:lanl-repo/fmt/3'. A look-up of this placeholder value in the sample DIP Table (Table 4) reveals that this placeholder value is associated with a service with Service Identifier 'info:lanl-repo/service/marc_2_mods', and that DIM code is available at 'http://purl.lanl.gov/dip/methods/marctomods.js'. Hence, this service must be added to the DIDL document. This requires:

- Inserting the actual DIM code as a `Component/Resource` construct in the DIDL document. Hereby, the Service Identifier 'info:lanl-repo/service/marc_2_mods' is also inserted as the identifier of the `Item` that contains the `Component/Resource`.
- Associating the inserted DIM with the `Component` which had 'info:lanl-repo/fmt/3' as the placeholder value. This is achieved by inserting an `objectType` element from the MPEG-21 DIP XML Namespace, and associating it with the DIM using the `ObjectType/Argument` technique of MPEG-21 DIP described above. As can be seen in Annex B, the association has been achieved through the value 'urn:uuid:8f64eabf'.

The result of applying the dynamic DIM Insertion process to the MARCXML metadata datastream of our sample Digital Object is shown in Annex B; all entities involved in the process are highlighted.

In order to embed all relevant DIMs into a DIDL document upon dissemination, the process explained above is repeated for every placeholder entry found in the DIDL document. The left part of Figure 5 provides a conceptual illustration of the process.

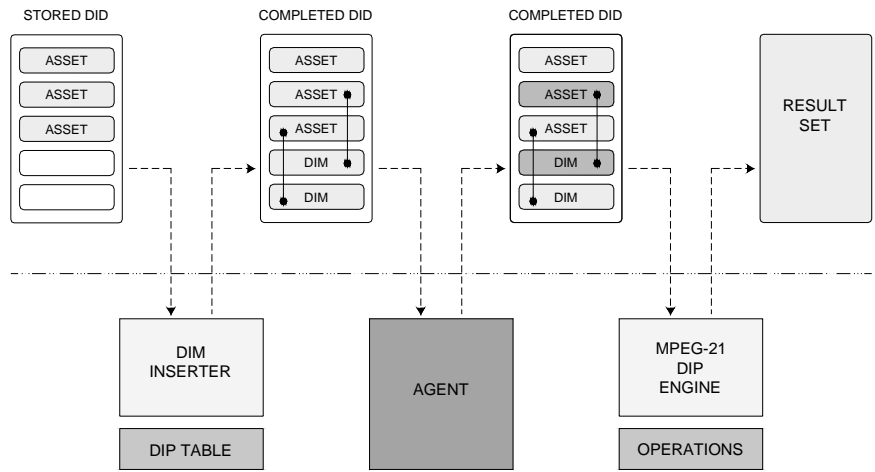

**Figure 5.** Dynamic dissemination of stored DIDL documents

### 5.3 Processing dissemination requests using the MPEG-21 DIP Engine

The MPEG-21 DIP Engine is capable of processing a Digital Object upon request of an agent. An MPEG-21 DIP Engine is able to respond to processing requests in which the following information is conveyed:

- An actual DIDL document
- An identification of the XML element of the DIDL document for which the service is requested
- An identification of the method – DIM – contained in the DIDL document that implements the requested service.

The actual functioning of the LANL MPEG-21 DIP Engine is illustrated in the right part of Figure 5. Upon receipt of a service request, the MPEG-21 DIP Engine extracts the identified XML element of the DIDL document, as well as the identified DIM. As was mentioned, DIMs are expressed using a language that normatively includes ECMAScript. This ECMAScript does not actually contain all the code required for the implementation of the service request. Rather, the ECMAScript is used by the MPEG-21 DIP Engine to bootstrap the process of implementing the service request and to actually coordinate it. To that end, the ECMAScript contains calls to so-called Digital Item Operations. Two types of Digital Item Operations are distinguished in MPEG-21 DIP:

- Digital Item Base Operations – DIBOs: DIBOs are Digital Item Operations that must be supported by every compliant MPEG-21 DIP Engine. Currently defined DIBOs include operations to manipulate DIDL documents at the XML level – these operations use the W3C DOM Core API [23] – and operations that have general application across a wide range of domains, applications and media types; some of which may require rights checking – e.g. `Play`, `Print`, and `Adapt`.
- Digital Item extension Operations – DIXOs: DIXOs are Digital Item Operations that are not defined by MPEG-21 DIP, but are rather defined at the level of specific communities, and applications. DIXOs are the extensibility mechanism built into the MPEG-21 DIP Engine.

An MPEG-21 DIP Engine is typically thought of as a client-side software component that operates on a user terminal. In aDORe, a server-side prototype implementation of an MPEG-21 DIP Engine has been developed. It currently supports only a few of the DIBOs defined by MPEG-21. But various DIXOs have been coded that implement services that are relevant for and specific to the nature of the collection stored in aDORe. As a matter of fact, all services listed in Table 3 are implemented using DIXOs. Also, Java-based templates have been developed to facilitate the straightforward creation of more DIXOs. For example, one template can be used as the basis for all DIXOs that use XSL transforms, another template for DIXOs that need to call external applications, and yet another one for the interaction with Web Services. aDORe is expected to be fully compatible with MPEG-21 Reference Software [7], and once the reference implementation of the MPEG-21 DIP Engine becomes available, it will be integrated in the aDORe environment to replace the current prototype implementation.

## 6. Accessing the aDORe environment: The OAI-PMH Federator and the OpenURL Resolver

In order to facilitate the retrieval of stored materials, two front-ends to aDORe are introduced. The front-ends hide the complexity of the aDORe environment to client-applications, and fulfill specific functions as summarized in Table 5:

- The *OAI-PMH Federator* provides an interface through which disseminations of stored DIDL documents can be requested. This interface is typically used by machines, and responses are expressed as XML. In OAIS terms, this interface facilitates the request of Dissemination Information Packages (OAIS DIPs). Available OAIS DIPs include the stored DIDL documents themselves as well as transformations thereof, such as the representation of a stored DIDL document using another complex object format than MPEG-21 DID, and the previously mentioned Completed DIDL document. The OAI-PMH Federator exposes the aDORe environment as a single OAI-PMH repository, in which the Package Identifiers (without XML ID fragment) of stored DIDL documents act as the OAI-PMH `identifier`. The OAI-PMH access provided by the Federator allows for batch retrieval (using the OAI-PMH `ListRecords` verb), as well as for access to an individual DIDL document (using the OAI-PMH `GetRecord` verb).

- The *OpenURL Resolver* provides an interface through which disseminations of Digital Objects and their constituent datastreams can be requested. This interface is provided for immediate presentation of the disseminations to end-users, and methods from which the disseminations result include 'show PDF' and 'play video'. In OAIS terms, this interface facilitates the request of OAIS Result Sets. The OpenURL Gateway provides an interface to aDORe that is compliant with the NISO OpenURL Standard, and in which both Content Identifiers and Package Identifiers (including the XML ID fragment) can be used to identify the requested resource.

| aDORe front-end | interface standard | identifier | OAIS access type | # items in response |
|---|---|---|---|---|
| OAI-PMH Federator | OAI-PMH | Package Identifier | OAIS DIP | 1 or more |
| OpenURL Resolver | NISO OpenURL | Content Identifier, Package Identifier (with XML ID fragment) | OAIS Result Set | 1 |

**Table 5.** Characteristic of the aDORe front-ends

## 6.1 An OAIS DIP interface to aDORe: The OAI-PMH Federator

The OAI-PMH Federator is introduced in the environment for the following reasons:

- As was described earlier, harvesters need to be aware of the structure of the aDORe environment in order to be able to collect DIDL documents. Required knowledge includes the existence of the Repository Index and the Identifier Locator, as well as the protocol to interface with them. This need for this prior know-how hinders the use of off-the-shelf OAI-PMH harvesting tools and requires the use of special-purpose clients.

- In addition to being able to disseminate stored DIDL documents from aDORe, there is a need for disseminations of various transformations thereof. For example, for reasons of interoperability, the dissemination of stored DIDL documents rendered according to complex object formats such as METS [25], IMS-CP [13], XFDU [6] is desirable. It is also attractive to be able to feed the Identifier Locator with the bare essentials of a stored DIDL document (the contained identifiers) rather than with the full-blown DIDL document itself. Also, there is a need to be able to request a Completed DIDL document with embedded Digital Item Methods. Supporting such transforms at the level of each Autonomous OAI-PMH Repository would significantly increase the complexity of the tools required for their implementation. Again, no off-the-shelf OAI-PMH repository tools could be used.

The OAI-PMH Federator addresses both problems:

- It exposes the whole aDORe environment as a single OAI-PMH repository, by translating incoming OAI-PMH requests into appropriate requests targeted at the Repository Index, Identifier Locator and Autonomous OAI-PMH Repositories. Since many of these components are themselves OAI-PMH repositories, the OAI-PMH Federator operates its private OAI-PMH harvester. Logic built into the OAI-PMH Federator ensures that the responses received from the various components are interpreted correctly, and, whenever appropriate, handed over to downstream harvesters as valid OAI-PMH responses.
- It allows for requesting transforms of stored DIDL documents by handing off transformation requests to the previously described MPEG-21 DIP Engine.

The OAI-PMH Federator is an OAI-PMH repository with the following characteristics:

- It has a unique, persistent `baseURL`, the http address `BaseURL(Federator)`.
- The `identifier` used by the OAI-PMH is the Package Identifier.
- The `datestamp` used by the OAI-PMH is the creation DateTime of the DIDL document.
- DIDL is the natively supported metadata format, but, through dynamic processing of DIDL documents by the MPEG-21 DIP Engine, potentially many other metadata formats can be supported. The term metadata format must be interpreted broadly, as the `metadataPrefix` argument in harvesting requests issued against the OAI-PMH Federator can be used to express several types of transformations that can be applied to stored DIDL documents:
  - Transformations that map DIDL to another complex object model such as IMS-CP. In this case, the value for the `metadataPrefix` argument in harvesting requests could be IMS-CP, and the IMS-CP XML Schema would define the metadata format.
  - Manipulations of stored DIDL documents, the result of which remains a DIDL document. An example is the request of Completed DIDL documents. In this case, the metadata format will remain DIDL, but the nature of the harvesting request will need to be further clarified through the `metadataPrefix`, i.e. DIDL:completed.
- The supported `granularity` is seconds-level.
- In order to support harvesting from selected Autonomous OAI-PMH Repositories, if this would be required, the OAI-PMH Federator can expose an OAI-PMH `set` structure according to their `baseURLs`. Other `set` structures implemented by all Autonomous OAI-PMH Repositories can also be exposed by the OAI-PMH Federator. In the current aDORe implementation a collection-oriented and format-oriented `set` structure is available.

## 6.2 The OAI-PMH Federator: A step-by-step example

The interaction of an off-the-shelf OAI-PMH harvester with the aDORe environment via the OAI-PMH Federator is illustrated by means of the following `GetRecord` requests:

```
[BaseURL(Federator)?
```

```
        verb=GetRecord&identifier=info:lanl-repo/i/58f202ac&
        metadataPrefix=DIDL
and
[BaseURL(Federator)?
        verb=GetRecord&identifier=info:lanl-repo/i/58f202ac&
        metadataPrefix=IMS-CP]
```

These are the steps involved in generating the appropriate response:

- Through interaction with the Identifier Locator, the OAI-PMH Federator finds out about the location of the DIDL document with Package Identifier 'info:lanl-repo/i/58f202ac', namely `BaseURL(3)`. If no entry for 'info:lanl-repo/i/58f202ac' exists in the Identifier Locator, the OAI-PMH Federator generates an `idDoesNotExist` error response.
- The OAI-PMH Federator obtains the stored DIDL document by issuing a `GetRecord` request `[BaseURL(3)?`
      `verb=GetRecord&identifier=info:lanl-repo/i/58f202ac&`
      `metadataPrefix=DIDL]`.
  - If the `metadataPrefix` requested in the original `GetRecord` request was DIDL, no special actions need to be undertaken.
  - If the `metadataPrefix` requested in the original `GetRecord` request was IMS-CP, via the DIP Table, the OAI-PMH Federator can determine whether the requested `metadataPrefix` (i.e. IMS-CP) is supported. If yes, the OAI-PMH Federator passes the DIDL document on to the DIM Inserter, which dynamically adds the transform that – in the DIP Table – corresponds to IMS-CP and associates it with the *container* entity of the DIDL document. Then, the OAI-PMH Federator calls the MPEG-21 DIP Engine to have it apply the transform. If the requested `metadataPrefix` is not supported, a `cannotDisseminateFormat` error response is generated.
- The OAI-PMH Federator embeds the record that results from the previous step in a correct OAI-PMH response. This may include inserting `set` membership information in the `headers` of the responses.

## 6.3 An OAIS Result Set interface to aDORe: The OpenURL Resolver

Through OAI-PMH harvesting, downstream applications obtain DIDL documents and use the represented Digital Objects and contained datastreams in the creation of their services. For example, a search engine might extract all textual information from harvested DIDL documents and make it available for searching by end-users. In this case, brief search results will point back to the corresponding resource stored in aDORe. Obviously, in such pointers the identification of the resource will play a crucial role, and both Package Identifiers (including the associated XML ID fragments), and Content Identifiers can be involved. Also, as was described, the aDORe environment allows delivery of various disseminations of stored resources through the application of services to those resources. Hence, requests to retrieve a resource should not only identify the requested resource but also the service that needs to be applied to it. The latter is equivalent to conveying the Service Identifier, as shown in Table 4, in the dissemination request.

The NISO OpenURL Standard [30] provides a perfect framework to convey such dissemination requests. The initial OpenURL specification [41] was specifically introduced for the purpose of reference linking, and was targeted at facilitating the provision of context-sensitive service links for popular types of scholarly works such as journal articles and books [39]. Hereby, identifiers and metadata describing the work are conveyed using a controlled-vocabulary HTTP GET request to a user-specific linking server, which uses a rules-based approach to provide a user with appropriate services pertaining to the work. A generalization of the essential components of the initial OpenURL solution [38] inspired the nature of the NISO OpenURL Standard [30], which provides a generic framework for the delivery of context-sensitive services pertaining to whichever type of resource referenced in a networked environment. To that end, the OpenURL Standard introduces the notion of a ContextObject, which is an information construct that contains descriptions of various entities involved in the process of providing context-sensitive services. The entities are shown in Table 6.

| Entity | Definition |
|---|---|
| Referent | The entity about which the ContextObject was created – the referenced resource |
| ReferringEntity | The entity that references the Referent |
| Requester | The entity that requests services pertaining to the Referent |
| ServiceType | The entity that defines the type of service requested |
| Resolver | The entity at which a request for services is targeted |
| Referrer | The entity that generated the ContextObject |

**Table 6.** The NISO OpenURL ContextObject

Each entity of the ContextObject can be described by means of identifiers, by means of metadata, and/or by means of private data. A ContextObject can be represented in many ways, and currently, a Key/Encoded-Value (KEV) representation and an XML representation have been defined and are registered in the OpenURL Registry (see: http://www.openurl.info/registry/) . A representation of a ContextObject can be transported to a networked system named a Resolver, in order to request services pertaining to the Referent described in it. To decide upon the nature of such services, the Resolver may take entities other than the Referent into account. The transport of a ContextObject can occur using various network protocols, and currently transport over HTTP and HTTPS have been defined.

From the above, two mappings follow easily:
- The resource stored in aDORe for which a dissemination is requested corresponds to the Referent of the OpenURL ContextObject.
- The service that needs to be applied to obtain the requested dissemination of the resource corresponds to the ServiceType of the OpenURL ContextObject.

An OpenURL Resolver is introduced in aDORe at `BaseURL(OpenURL Resolver)` to which all requests for the dissemination of stored Digital Objects or contained datastreams are targeted. Using examples provided earlier in this paper (Tables 1-3), the HTTP transport of the OpenURL Standard, and its KEV representation of the ContextObject, the following are valid OpenURLs to convey dissemination requests to this OpenURL Resolver:

- Display a Table of Contents for our sample Digital Object with Content Identifier 'info:doi/10.123/44455':
  ```
  [BaseURL(OpenURL Resolver)?
      url_ver=Z39.88-2004 &
      rft_id=info:doi/10.123/44455 &
      svc_id=info:lanl-repo/service/]
  ```
  Using Package Identifiers, this request can also be formulated as:
  ```
  [BaseURL(OpenURL Resolver)?
      url_ver=Z39.88-2004 &
      rft_id=info:lanl-repo/i/58f202ac#uuid-00005e90 &
      svc_id=info:lanl-repo/service/]
  ```
- Display the MARCXML metadata record with Content Identifier 'info:pmid/2225887' that is a constituent datastream of our sample Digital Object as a MODS [26] metadata record:

```
[BaseURL(OpenURL Resolver)?
    url_ver=Z39.88-2004 &
    rft_id=info:pmid/2225887 &
    svc_id=info:lanl-repo/service/marc_2_mods]
```
   Again, this request could be formulated in terms of Package Identifiers.
- Display the PDF article that is a constituent datastream of our sample Digital Object. As no Content Identifier exists for this datastream, the request must be formulated using Package Identifiers. Also, as display of the PDF "as is" is requested, no ServiceType is being conveyed in the request:
```
[BaseURL(OpenURL Resolver)?
    url_ver=Z39.88-2004 &
    rft_id=info:lanl-repo/i/58f202ac#uuid-00004a42
```

Currently, the Referent and ServiceType are the only entities of the ContextObject that are used in aDORe OpenURL requests. But, NISO OpenURL allows expressing other entities that can be taken into account to deliver context-sensitive dissemination requests. For example, conveying Requester information may be of particular interest, as this would allow adapting the actual dissemination to the agent requesting it. Requester information could convey identity, and this would allow responding differently to the same service request depending on whether the requesting agent is human or machine. Or different humans could receive different disseminations based on recorded preferences or access rights. But the NISO OpenURL specification is purposely very generic and extensible, and would also support to convey the characteristics of a user's terminal, and/or the user's location via the Requester entity. Such information could be passed on to the MPEG-21 DIP Engine, and be taken into account in the delivery of an actual dissemination. As a matter of fact, the expressiveness of NISO OpenURL seems to resonate nicely with the nature of the MPEG-21 Digital Item Adaptation (DIA) [19] effort that focuses on the description of usage environment characteristics such as terminal capabilities, network conditions and other contextual information.

## 6.4 The OpenURL Resolver: a step-by-step example

This section provides a walkthrough of the processes involved in responding to the previously described OpenURL request:
```
[BaseURL(OpenURL Resolver)?
    url_ver=Z39.88-2004 &
    rft_id=info:pmid/2225887 &
    svc_id=info:lanl-repo/service/marc_2_mods]
```

- Through interaction with the Identifier Locator, the OpenURL Resolver learns that the Referent with Content Identifier 'info:pmid/2225887' has the corresponding Package Identifier 'info:lanl-repo/i/58f202ac#uuid-8881b35e'. This identifies a DIDL element contained in the DIDL document with Package Identifier 'info:lanl-repo/i/58f202ac', which is located in the OAI-PMH repository with baseURL 'BaseURL(3)'.
- The OpenURL Resolver retrieves this DIDL document from its OAI-PMH repository by issuing the OAI-PMH GetRecord request
```
[BaseURL(3)?
    verb=GetRecord&
    identifier=info:lanl-repo/i/58f202ac&
    metadataPrefix=DIDL].
```
- The OpenURL Resolver passes the retrieved DIDL document on to the DIM Inserter, which dynamically adds DIMs to the document based on contained 'placeholder' values and look-ups in the DIP Table.
- The DIM Inserter returns the Completed DIDL document (see Annex B) to the OpenURL Resolver.

- The OpenURL Resolver passes the Completed DIDL document, the Package Identifier of the Referent, and the identifier of the ServiceType to the MPEG-21 DIP Engine.
- Through inspection of the DIDL document, the MPEG-21 DIP Engine determines whether the requested service with identifier 'info:lanl-repo/service/marc_2_mods' can at all be applied to the element with XML ID 'uuid-8881b35e'. This is achieved by checking the `ObjectType/Argument` correspondence.
- Assuming that the service can indeed be applied, the appropriate DIM is extracted from the DIDL document, and executed by the MPEG-21 DIP Engine. This DIM calls a DIXO that transforms the content of the element with XML ID 'uuid-8881b35e' from MARCXML to MODS.
- The MPEG-21 DIP Engine returns the MODS record to the OpenURL Resolver, which delivers it to the requesting agent.

## 7. Conclusion

This paper has described the aDORe Digital Object Repository architecture as it has been defined over the past 2 years by the Digital Library Research & Prototyping Team of the Research Library of the Los Alamos National Laboratory. Some core design choices in the aDORe environment give it the following attractive characteristics:

- The modular nature of the design, especially the ability to store DIDL documents in multiple Autonomous OAI-PMH Repositories provides reassurances regarding the scalability of the solution.
- Handling the binding of dissemination methods to stored Digital Objects in a dynamic manner avoids what would likely become a significant workload in editing stored DIDL documents to update embedded methods.
- Avoiding editing of stored DIDL documents is further realized by the creation of a new DIDL document whenever a new version of a previously ingested Digital Object needs to be ingested. As all such versions share the same Content Identifier, the parallel identification mechanism (Content Identifier, Package Identifier) ensures that the different versions remain distinguishable. And, through the Identifier Locator, all versions can be located. The parallel identification approach also allows addressing resources that have no Content Identifier.
- The standards-based nature of the design allows the use of off-the-shelf, open source tools in all areas of the aDORe environment. This has resulted in significant time savings for the development, but also allows for inspection, improvement, modification of code if required. For example, various free and open-source OAI-PMH tools are used, including OAICat [32], OAIHarvester [33], and OAI Viewer [31]. Also, it is expected that reference implementations of an OpenURL parser, an MPEG-21 DIDL parser and the MPEG-21 DIP Engine will become available for integration into the environment.
- The modular and protocol-based nature of the design suggests the possibility of an implementation of the aDORe architecture in a distributed Web environment, even though it was conceived and implemented for a local environment. A particular area of interest is a federation of Institutional Repositories [20]. Furthermore, the protocol-based design provides the flexibility to choose from or migrate between different software implementations for aDORe components, without the overall design and functionality being affected. Such migration is already underway in the context of the creation of a more robust implementation.
- The design introduces a novel use of the NISO OpenURL Standard which begs reflection on its use as an interoperable interface to request disseminations of resources stored in heterogeneous repositories. The fact that OpenURL allows to convey information that is relevant for the delivery of a context-sensitive response to a dissemination request makes the idea all the more appealing.

- As far as can be verified, at the time of writing, aDORe is the first environment in which various MPEG-21 technologies are used at a significant scale. If nothing else, aDORe strongly suggests the significance of the MPEG-21 standardization effort to the Digital Library, Archiving, and Learning Object communities. Furthermore, aDORe has illustrated the feasibility of an MPEG-21 DIP Engine as a server-side component, thereby adjusting the traditional MPEG-21 perspective of it being a software component operating on a user terminal such as a PDA. Obviously, when such terminal-based tools become available, aDORe will be able to directly communicate with them.

An initial aDORe implementation has been brought in production that currently stores about 30,000,000 DIDL documents, a figure that is expected to at least double in the next year. The current aDORe implementation is developed in Java and Perl and has been tested in Linux and Solaris environments. Off-the-shelf software packages have been used throughout the implementation; these include utilities to write Internet Archive ARC files from the Danish netarchive project, standard XML tools to write XMLTapes, and Berkeley DB to index them to enable OAI-PMH access. Both the OAI-PMH Federator and the Repository Index are based on OCLC's OAICat, while OAI-PMH harvesters used by downstream applications are based on OCLC's OAIHarvester. The modular and protocol-based nature of the design makes it possible to distribute aDORe components across multiple machines; as a matter of fact, the Autonomous OAI-PMH Repositories, the OAI-PMH Federator, the OpenURL Resolver, the Repository Index, and the Identifier Locator can all be running on separate machines.

The performance and scalability of the Identifier Locator is critical to the aDORe environment. The current implementation is based on a single MySQL database which has the critical information loaded into RAM. This implementation has worked successfully with the current amount of stored Digital Objects. However it is expected that optimizations will be required as the number of stored Digital objects further increases, and as the environment is more intensely accessed. Anticipated optimizations include distributing the data handled by the Identifier Locator across multiple MySQL servers running on a blade environment.

Various downstream applications harvest on a recurrent basis from aDORe. Two harvesters work on behalf of search engines (Verity, Lucene) that support discovery of ingested materials, and another on behalf of an application aimed at dynamically de-duplicating bibliographic information which is built on top of the Netrics fuzzy search engine. In addition to the ongoing ingestion of scholarly assets obtained from primary and secondary publishers, work is under way aimed at ingesting materials as diverse as technical reports authored by LANL employees, datasets, videotaped presentations, and materials gathered through focused Web crawling. Also, the implementation of more robust versions for several of the described components is underway, and consideration is being given to share the aDORe code-base with interested parties.

## Acknowledgments

The authors would like to thank their previous team members Patrick Hochstenbach (now at Ghent University) and Henry Jerez (now at CNRI) for their contributions to the reported work. Many thanks also to Rick Luce, the director of the LANL Research Library, for his ongoing support for the reported work; and to Miriam Blake, Mariella Di Giacomo and Beth Goldsmith (LANL Research Library Development Team) for their efforts in deploying the aDORe environment. Thanks to Michael L. Nelson for proofreading a draft of this paper.

Jeroen Bekaert also wishes to thank the Fund for Scientific Research (Flanders, Belgium) for his Ph.D. scholarship. The reported research is partly funded by a grant from the Library of Congress's National Digital Information Infrastructure and Preservation Program.

## Annex A: A DIDL document representing the sample Digital Object

```xml
<?xml version="1.0" encoding="UTF-8"?>
<!-- DIDL document is Archival Information Package -->
<!-- Package Identifier provided as value of DIDid attribute -->
<didl:DIDL diext:DIDid="info:lanl-repo/i/58f202ac"
   diext:DIDcreated="2004-06-22T18:07:18Z" xmlns:didl="urn:mpeg:mpeg21:2002:02-DIDL-NS"
   xmlns:diext="http://library.lanl.gov/2004-04/STB-RL/DIEXT"
   xmlns:xsi="http://www.w3.org/2001/XMLSchema-instance">
   <!-- Container element representing a container entity -->
   <didl:Container id="uuid-73d2247e">
      <!-- Container-level Placeholder -->
      <didl:Descriptor>
         <didl:Statement mimeType="text/xml; charset=UTF-8">
            <diadm:Admin xmlns:diadm="http://library.lanl.gov/2004-01/STB-RL/DIADM">
               <dc:format xmlns:dc="http://purl.org/dc/elements/1.1/">info:lanl-repo/pro/DIDL</dc:format>
            </diadm:Admin>
         </didl:Statement>
      </didl:Descriptor>
      <!-- Top-level Item representing the Digital Object -->
      <didl:Item id="uuid-00005e90">
         <!-- Content Identifier of the Digital Object -->
         <didl:Descriptor>
            <didl:Statement mimeType="text/xml; charset=UTF-8">
               <dii:Identifier xmlns:dii="urn:mpeg:mpeg21:2002:01-DII-NS">
                  info:doi/10.123/44455</dii:Identifier>
            </didl:Statement>
         </didl:Descriptor>
         <!-- Item-level Placeholder -->
         <didl:Descriptor>
            <didl:Statement mimeType="text/xml; charset=UTF-8">
               <diadm:Admin xmlns:diadm="http://library.lanl.gov/2004-01/STB-RL/DIADM">
                  <dc:format xmlns:dc="http://purl.org/dc/elements/1.1/">info:lanl-repo/pro/paper</dc:format>
               </diadm:Admin>
```

```
            </didl:Statement>
        </didl:Descriptor>
    <!-- Sub-Item containing a MARCXML metadata record -->
    <didl:Item id="uuid-8881b35e">
        <!-- Content Identifier of the MARCXML metadata record -->
        <didl:Descriptor>
            <didl:Statement mimeType="text/xml; charset=UTF-8">
                <dii:Identifier xmlns:dii="urn:mpeg:mpeg21:2002:01-DII-NS">
                    info:pmid/2225887</dii:Identifier>
            </didl:Statement>
        </didl:Descriptor>
        <!-- Sub-Item-level Placeholder -->
        <didl:Descriptor>
            <didl:Statement mimeType="text/xml; charset=UTF-8">
                <diadm:Admin xmlns:diadm="http://library.lanl.gov/2004-01/STB-RL/DIADM">
                    <dc:format xmlns:dc="http://purl.org/dc/elements/1.1/">
                        info:lanl-repo/pro/metadata</dc:format>
                </diadm:Admin>
            </didl:Statement>
        </didl:Descriptor>
        <!-- Component containing the MARCXML datastream -->
        <didl:Component id="uuid-0000a01c">
            <!-- Component-level Placeholder / Format -->
            <didl:Descriptor>
                <didl:Statement mimeType="text/xml; charset=UTF-8">
                    <diadm:Admin xmlns:diadm="http://library.lanl.gov/2004-01/STB-RL/DIADM">
                        <dc:format xmlns:dc="http://purl.org/dc/elements/1.1/">
                            info:lanl-repo/fmt/3</dc:format>
                    </diadm:Admin>
                </didl:Statement>
            </didl:Descriptor>
            <!-- The actual MARCXML datastream -->
            <didl:Resource mimeType="text/xml; charset=UTF-8">
                <record xmlns="http://www.loc.gov/MARC21/slim">
                    <leader>01748cam 220036101 45Y0</leader>
                    <controlfield tag="001">LANLb10012252</controlfield>
                    <controlfield tag="003">LANL</controlfield>
                    <controlfield tag="005">20030527112640.0</controlfield>
                    <controlfield tag="008">840202s1983  nmua   tb  00010 eng d</controlfield>
                    <datafield tag="035" ind1=" " ind2=" ">
                        <subfield code="a">GLIS00012252</subfield>
                    </datafield>
                    ...
                </record>
            </didl:Resource>
        </didl:Component>
    </didl:Item>
    <!-- Component containing the PDF paper -->
    <didl:Component id="uuid-00004a42">
        <didl:Descriptor>
            <!-- Component-level Placeholder / Format -->
            <didl:Descriptor>
                <didl:Statement mimeType="text/xml; charset=UTF-8">
                    <diadm:Admin xmlns:diadm="http://library.lanl.gov/2004-01/STB-RL/DIADM">
                        <dc:format xmlns:dc="http://purl.org/dc/elements/1.1/">
                            info:lanl-repo/fmt/5</dc:format>
                    </diadm:Admin>
                </didl:Statement>
            </didl:Descriptor>
            <!-- The actual PDF datastream -->
            <didl:Resource mimeType="application/pdf" encoding="base64">
PSJjIj5jMTk5My48L3N1YmZpZWxkPg0KICAgIDw9uIHhtbG5sZSJodHgKICAgIDxk
dGFnPSIzMDAiIGluZDE9IiAiIGluZDI9IiI9IiIgICAgICAgICAgPHN1YmZpZWxkIGNv
cmVzdG9yZWQgdG8g...</didl:Resource>
        </didl:Component>
    </didl:Item>
    </didl:Container>
</didl:DIDL>
```

## Annex B: The DIDL document from Annex A containing a DIM associated with an MPEG-21 DID *component*

```
<?xml version="1.0" encoding="UTF-8"?>
<!-- DIDL document is Archival Information Package -->
<!-- Package Identifier provided as value of DIDid attribute -->
<didl:DIDL diext:DIDid="info:lanl-repo/i/58f202ac"
    diext:DIDcreated="2004-06-22T18:07:182" xmlns:didl="urn:mpeg:mpeg21:2002:02-DIDL-NS"
    xmlns:diext="http://library.lanl.gov/2004-04/STB-RL/DIEXT"
```

```xml
  xmlns:xsi="http://www.w3.org/2001/XMLSchema-instance">
  <!-- Container element representing a container entity -->
  <didl:Container id="uuid-73d2247e">
    <!-- Container-level Placeholder -->
    <didl:Descriptor>
      <didl:Statement mimeType="text/xml; charset=UTF-8">
        <diadm:Admin xmlns:diadm="http://library.lanl.gov/2004-01/STB-RL/DIADM">
          <dc:format xmlns:dc="http://purl.org/dc/elements/1.1/">info:lanl-repo/pro/DIDL</dc:format>
        </diadm:Admin>
      </didl:Statement>
    </didl:Descriptor>
    <!-- Top-level Item representing the Digital Object -->
    <didl:Item id="uuid-00005e90">
      <!-- Content Identifier of the Digital Object -->
      <didl:Descriptor>
        <didl:Statement mimeType="text/xml; charset=UTF-8">
          <dii:Identifier xmlns:dii="urn:mpeg:mpeg21:2002:01-DII-NS">
            info:doi/10.123/44455</dii:Identifier>
        </didl:Statement>
      </didl:Descriptor>
    <!-- Item-level Placeholder -->
    <didl:Descriptor>
      <didl:Statement mimeType="text/xml; charset=UTF-8">
        <diadm:Admin xmlns:diadm="http://library.lanl.gov/2004-01/STB-RL/DIADM">
          <dc:format xmlns:dc="http://purl.org/dc/elements/1.1/">info:lanl-repo/pro/paper</dc:format>
        </diadm:Admin>
      </didl:Statement>
    </didl:Descriptor>
    <!- Sub-Item containing a MARCXML metadata record -->
    <didl:Item id="uuid-8881b35e">
      <!-- Content Identifier of the MARCXML metadata record -->
      <didl:Descriptor>
        <didl:Statement mimeType="text/xml; charset=UTF-8">
          <dii:Identifier xmlns:dii="urn:mpeg:mpeg21:2002:01-DII-NS">
            info:pmid/2225887</dii:Identifier>
        </didl:Statement>
      </didl:Descriptor>
      <!-- Sub-Item-level Placeholder -->
      <didl:Descriptor>
        <didl:Statement mimeType="text/xml; charset=UTF-8">
          <diadm:Admin xmlns:diadm="http://library.lanl.gov/2004-01/STB-RL/DIADM">
            <dc:format xmlns:dc="http://purl.org/dc/elements/1.1/">
              info:lanl-repo/pro/metadata</dc:format>
          </diadm:Admin>
        </didl:Statement>
      </didl:Descriptor>
      ...
      <!-- Component containing the MARCXML datastream -->
      <didl:Component id="uuid-0000a01c">
        <!-- Component-level Placeholder / Format -->
        <didl:Descriptor>
          <didl:Statement mimeType="text/xml; charset=UTF-8">
            <diadm:Admin xmlns:diadm="http://library.lanl.gov/2004-01/STB-RL/DIADM">
              <dc:format xmlns:dc="http://purl.org/dc/elements/1.1/">
                info:lanl-repo/fmt/3</dc:format>
            </diadm:Admin>
          </didl:Statement>
        </didl:Descriptor>
        <!-- ObjectType of the MARCXML datastream, added by DIM Inserter -->
        <!-- Corresponds with Argument of DIM (MARCXML to MODS) below -->
        <didl:Descriptor>
          <didl:Statement mimeType="text/xml; charset=UTF-8">
            <dip:ObjectType xmlns:dip="urn:mpeg:mpeg21:2002:01-DIP-NS">
              urn:uuid:8f64eabf</dip:ObjectType>
          </didl:Statement>
        </didl:Descriptor>
        ...
        <!-- The actual MARCXML datastream -->
        <didl:Resource mimeType="text/xml; charset=UTF-8">
          <record xmlns="http://www.loc.gov/MARC21/slim">
            <leader>01748cam 220036101 45Y0</leader>
            <controlfield tag="001">LANLbl10012252</controlfield>
            <controlfield tag="003">LANL</controlfield>
            <controlfield tag="005">20030527112640.0</controlfield>
            <controlfield tag="008">840202s1983 nmua tb 00010 eng d</controlfield>
            <datafield tag="035" ind1=" " ind2=" ">
              <subfield code="a">GLIS00012252</subfield>
            </datafield>
            ...
          </record>
        </didl:Resource>
      </didl:Component>
    </didl:Item>
```

```
            <!-- Component containing the PDF paper -->
            <didl:Component id="uuid-00004a42">
               </didl:Descriptor>
                  <!-- Component-level Placeholder / Format -->
                  <didl:Statement mimeType="text/xml; charset=UTF-8">
                     <diadm:Admin xmlns:diadm="http://library.lanl.gov/2004-01/STB-RL/DIADM">
                        <dc:format xmlns:dc="http://purl.org/dc/elements/1.1/">
                           info:lanl-repo/fmt/5</dc:format>
                     </diadm:Admin>
                  </didl:Statement>
               </didl:Descriptor>
               ...
               <!-- The actual PDF datastream -->
               <didl:Resource mimeType="application/pdf" encoding="base64">
               PSJjIj5jMTk5My48L3N1YmZpZWxkPg0KICAgIDw9uIHhtbG5zSJodHgKICAgIDxk
               dGFnPSIzMDAiAiIGluZDE9IiAiIGluZDI9IiIOAIS AIPg0KICAgICAgPHN1YmZpZWxkIGNv
               cmVzdG9yZWQgdG8g...</didl:Resource>
            </didl:Component>
         </didl:Item>
         <!-- Item containing the DIM that implements the MARCXML to MODS service -->
         <!-- Inserted by the DIM Inserter after lookup of Placeholder/Format value info:lanl-repo/fmt/3 in
         the DIP Table -->
         <didl:Item id="uuid-6b479d14">
            <!-- Content Identifier of the DIM -->
            <didl:Descriptor>
               <didl:Statement mimeType="text/xml; charset=UTF-8">
                  <dii:Identifier xmlns:dii="urn:mpeg:mpeg21:2002:01-DII-NS">
                     info:lanl-repo/service/marc_2_mods</dii:Identifier>
               </didl:Statement>
            </didl:Descriptor>
            <didl:Component id="uuid-3886920b">
               <!-- Argument of the DIM. Corresponds with ObjectType attached to MARCXML datastream -->
               <didl:Descriptor>
                  <didl:Statement mimeType="text/xml; charset=UTF-8">
                     <dip:MethodInfo xmlns:dip="urn:mpeg:mpeg21:2002:01-DIP-NS">
                        <dip:Argument>urn:uuid:8f64eabf</dip:Argument>
                     </dip:MethodInfo>
                  </didl:Statement>
               </didl:Descriptor>
               <!-- DIM ECMAScript -->
               <didl:Resource mimeType="application/mp21-method"
                  ref="http://purl.lanl.gov/dip/methods/marctomods.js"/>
            </didl:Component>
         </didl:Item>
      </didl:Container>
   </didl:DIDL>
```